\documentclass[summary]{ursi_sdb}

\title{LuSEE 'Night':  The Lunar Surface Electromagnetics Experiment}

\author{Stuart D. Bale\affref{ref1}\affref{ref2}, Neil Bassett\affref{ref3}, Jack O. Burns\affref{ref3}, Johnny Dorigo Jones\affref{ref3}, Keith Goetz\affref{ref4}, Christian Hellum-Bye\affref{ref5}, Sven Herrmann\affref{ref6}, Joshua Hibbard\affref{ref3}, Milan Maksimovic\affref{ref7}, Ryan McLean\affref{ref2},
Raul Monsalve\affref{ref2}\affref{ref8}\affref{ref9}, Paul O'Connor\affref{ref6}, Aaron Parsons\affref{ref5}, Marc Pulupa\affref{ref2}, Rugved Pund\affref{ref10}, David Rapetti\affref{ref3}\affref{ref11}\affref{ref12}, Kaja M. Rotermund\affref{ref13}, Ben Saliwanchik\affref{ref6}, An\v{z}e Slosar\affref{ref14}, David Sundkvist\affref{ref2}, and Aritoki Suzuki\affref{ref13}}

\affiliation{%
  \aff{ref1}{Physics Department, University of California, Berkeley, USA; e-mail: bale@berkeley.edu}
  \aff{ref2}{Space Sciences Laboratory, University of California, Berkeley, USA}
  \aff{ref3}{Center for Astrophysics \& Space Astronomy, University of Colorado, Boulder, CO, USA}
  \aff{ref4}{School of Physics and Astronomy, University of Minnesota, Minneapolis, USA}
  \aff{ref5}{Department of Astronomy, University of California, Berkeley, USA}
 \aff{ref6}{Instrumentation Division, Brookhaven National Laboratory, Upton, NY, USA}
\aff{ref7}{LESIA, Observatoire de Paris, Universit\'e PSL, CNRS, Sorbonne Universit\'e, Universit\'e de Paris, Meudon, France}
\aff{ref8}{Facultad de Ingenier\'ia, Universidad Cat\'olica de la Sant\'isima Concepci\'on, Concepci\'on, Chile}
\aff{ref9}{School of Earth and Space Exploration, Arizona State University, Tempe, USA}
\aff{ref10}{Department of Physics and Astronomy, Stony Brook University, Stony Brook, NY, USA}
\aff{ref11}{NASA Ames Research Center, Moffett Field, CA, USA}
\aff{ref12}{Research Institute for Advanced Computer Science, Universities Space Research Association, Washington, DC, USA}
\aff{ref13}{Lawrence Berkeley National Laboratory, Berkeley, CA, USA}
\aff{ref14}{Physics Department, Brookhaven National Laboratory, Upton NY, USA}
}

\begin{document}

\maketitle
\begin{abstract}
The Lunar Surface Electromagnetics Explorer 'LuSEE Night' is a low frequency radio astronomy experiment that will be delivered to the farside of the Moon by the NASA Commercial  Lunar Payload Services (CLPS) program in late 2025 or early 2026.  The payload system is being developed jointly by NASA and the US Department of Energy (DOE) and consists of a 4 channel, 50 MHz Nyquist baseband receiver system and 2 orthogonal $\sim$6m tip-to-tip electric dipole antennas.  LuSEE Night will enjoy standalone operations through the lunar night, without the electromagnetic interference (EMI) of an operating lander system and antipodal to our noisy home planet.  
\end{abstract}

\section{Science Motivation}
The radio sky below 20 MHz is not well-studied, primarily due to the $\mathcal{O}$(10 MHz) plasma frequency of the F-layer peak of the terrestrial ionosphere \cite{2022RvGeo..6000792B}, below which the ionosphere becomes opaque, but also due to the presence of intense shortwave radio transmissions \cite{1996GeoRL..23.1287K} and military over-the-horizon (A-band) radar. 

At these frequencies, the ambient sky is dominated by the synchrotron spectrum of cosmic ray electrons spiraling in the galactic magnetic field \cite{1978ApJ...221..114N,2008MNRAS.388..247D,2022A&A...668A.127P}.  This galactic synchrotron spectrum is brighter in the galactic plane above a few MHz  \cite{2001A&A...372..663M,2022A&A...668A.127P} but becomes attenuated in the plane by free-free absorption at lower frequencies \cite{2002ApJ...575..217P,2021ApJ...914..128C}.  Indeed, recent measurements by the Parker Solar Probe/FIELDS instrument \cite{2016SSRv..204...49B,2017JGRA..122.2836P} suggest that modifications to the state-of-the-art low frequency sky models \cite{2021ApJ...914..128C} are required in the form of small-scale angular structure, hemispheric asymmetries, and modifications to the galactic electron density map \cite{bassett2023,2022A&A...668A.127P}.

The brightest discrete radio sources below 50 MHz are likely to be the Sun \cite{2022ApJ...924...58S} and Jupiter \cite{2013P&SS...77....3P}; however there are other known astrophysical sources Cas A, Cyg A \cite{2009AJ....138..838H,2020A&A...635A.150D} present and plenty of 'discovery space'.  Nonthermal solar and Jovian radio sources will be dynamic and in some cases strongly polarized.  The known positions of the Sun, Jupiter and other sources will allow occultation studies as they set below the lunar horizon, potentially allowing studies of thermal emission from the extended solar corona. 

Characterizing the global spectrum of this low frequency foreground is a first step in a measurement strategy to access the high redshift ($z \sim$ 100) 21 cm signature associated with the so-called 'Dark Ages' \cite{2004PhRvL..92u1301L,2006PhR...433..181F,2021PSJ.....2...44B}.  The Dark Ages signal arose as the universe recombined and the cosmic microwave background (CMB) interacted with neutral hydrogen via the 21 cm spin-flip transition.  In principle, this absorption feature should be dominated by the thermal physics of the early universe -  any striking deviations may represent new physics.  However, the Dark Ages spectral feature is hidden more than 5 orders of magnitude below the galactic foreground such that any spectral structure (real and/or system chromaticity) in the foreground measurement will make recovery of the cosmological signal very challenging.  The LuSEE Night measurements will be an important step in this longer-term program.

\section{LuSEE Night System}
The LuSEE payload was selected by NASA in 2019 under the Lunar Surface Instrument and Technology Payloads (LSITP) program, which is itself in support of the Commercial Lunar Payload Services (CLPS) program.  As proposed, LuSEE  was largely heritage instrumentation from the 'FIELDS' experiment \cite{2016SSRv..204...49B} on the NASA Parker Solar Probe mission, comprised of magnetometers, electric field, and low frequency radio and plasma waves measurements.  In early 2020, NASA entered into a partnership with the DOE to promote LuSEE to a low frequency radio pathfinder and the LuSEE Night payload was separated from the original payload (now called 'LuSEE Lite' and destined for the Schr\"odinger Basin near the lunar South Pole, in early 2025).

LuSEE Night is designed to make low frequency ($\leq$ 50 MHz) full-Stokes, spectral density measurements of the radio sky in the pristine radio environment of the lunar farside.  To that end, LuSEE Night is required to operate continuously through the lunar night ($\sim$14 Earth days) with the Sun below the lunar horizon.  Furthermore, LuSEE Night requires that the CLPS lunar lander spacecraft {\em cease all operations} before the first nightfall and remain powered off for the duration of the mission.  This gives LuSEE Night full control of the electromagnetic environment and eliminates the risk of spacecraft-generated EMI (which bedeviled the Chang'e 4 radio astronomy experiment \cite{2019EPSC...13..529S}).

\begin{figure}[htbp]
  \centering
  \includegraphics[width=85mm]{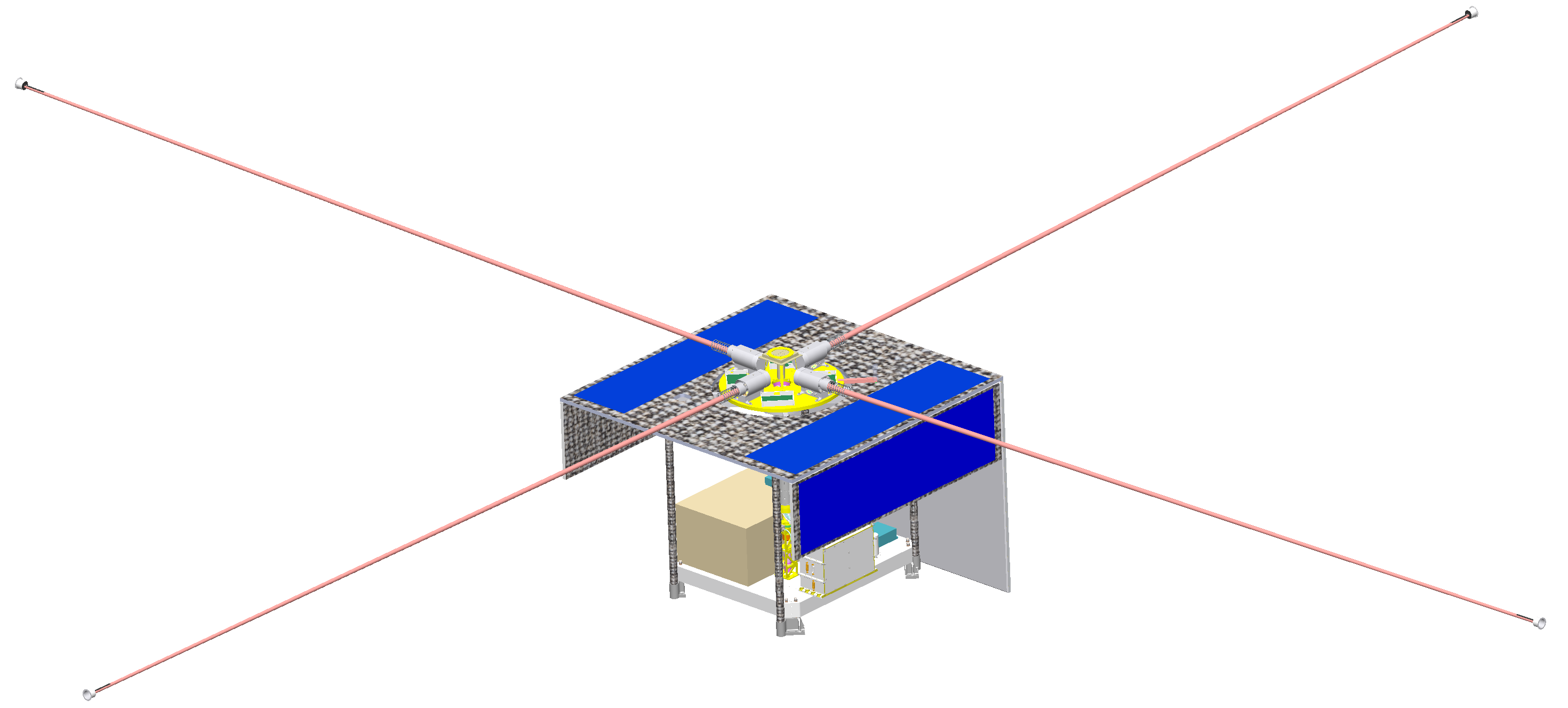}
  \caption{A CAD drawing of the LuSEE Night system showing the dipole antennas on the carousel, solar arrays (blue) and the large battery (gold box below).  The dipole antennas are $\sim$6m tip-to-tip, the upper deck is approximately 1m x 1m and the system is $\sim$70 cm tall, but will be mounted atop the lander.}
  \label{fig:system}
\end{figure}

Figure \ref{fig:system} is a CAD drawing of the LuSEE Night payload system showing the dipole antennas fully deployed.  LuSEE Night uses 3m BeCu stacer antenna elements \cite{2008SSRv..136..529B} configured in orthogonal, co-linear $\sim$6m tip-to-tip dipole pairs; a stacer is a cold-rolled, helical pitch spring that deploys by its own stored spring energy and forms a rigid tube of roughly 1 cm diameter.  The LuSEE Night antennas are mounted on a motor-driven carousel so that the antenna orientation can be rotated in the plane of the lunar surface.  This allows for a degree of freedom that can be used to understand the electrical coupling of the antenna system to the lunar lander structure and to the dielectric regolith below \cite{2018Icar..314..389G}.  The antenna carousel is only operable during the lunar day, when LuSEE Night solar arrays are illuminated and the system is power-positive.

The LuSEE Night structure occupies a cube (1m x 1m by 70cm high).  The top surface will support the four electric monopoles on a rotating platform as well as horizontal solar cells.  In addition the East and West sides of the cube will also include some vertical solar cells to improve the charging at dawn and dusk.  The LuSEE electronics package will be located within the cube and will be thermally isolated such that self-heating from the nighttime operations power will keep the electronics payload warm during the night.  One side of the cube will include south-facing radiator panels equipped with Parabolic Reflector Radiators (PRR ) developed at JPL \cite{bugby} to help reject day-time heating.

\begin{figure}[htbp]
  \centering
  \includegraphics[width=85mm]{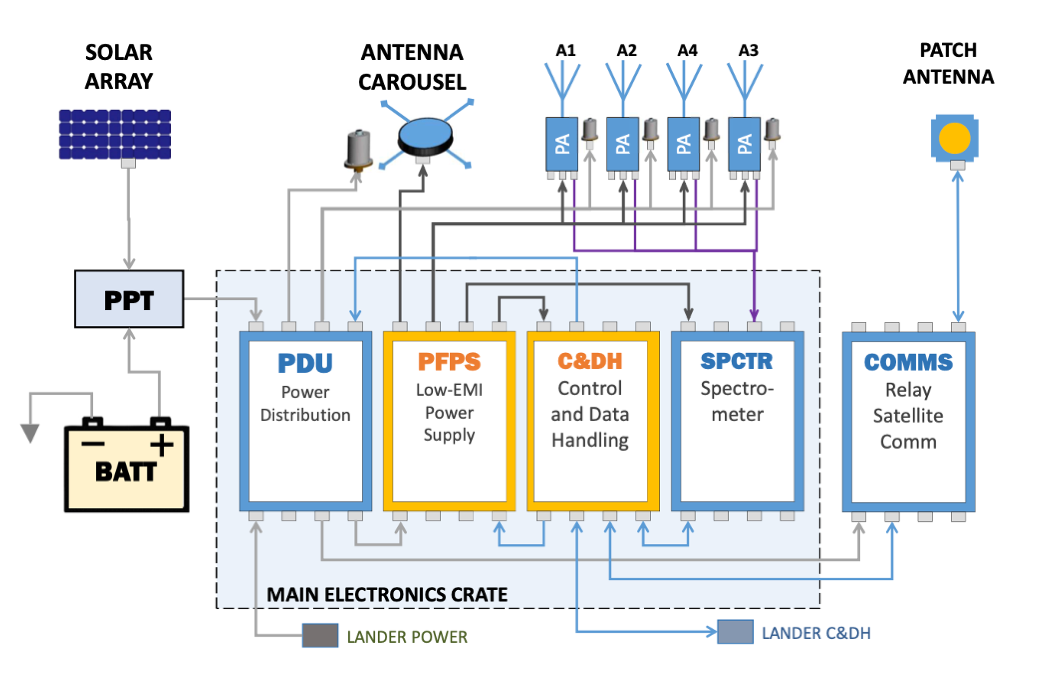}
  \caption{A block diagram of the LuSEE Night payload showing the instrument components - spectrometer, antennas, preamps, C\&DH computer, and power systems.  The LuSEE Night computer also operates the S-band communications system and motor drive to reorient the antenna carousel during daytime.}
  \label{fig:block}
\end{figure}

An instrument block diagram is shown in Figure \ref{fig:block}.  LuSEE Night will make a {\em voltage} measurement using a high impedance, low-noise JFET front end on each monopole antenna.  While a voltage measurement provides less gain than a balanced system, it can be more robust against impedance mismatch and internal reflections, which are difficult to correct after launch and landing. 

A spectrometer samples the four single-ended antenna voltages at 102.4 Msamples/sec and uses an FPGA to process the waveforms into auto- and cross-correlation spectra.  Stokes parameters can be computed using only cross-correlation products if desired, avoiding the antenna shotnoise present in the autocorrelations.  The spectrometer design is based on the Radio Frequency Spectrometer (RFS) system \cite{2017JGRA..122.2836P} from the Parker Solar Probe/FIELDS instrument \cite{2016SSRv..204...49B}.  A DC-DC power converter system will operate at well-defined, stable (crystal-controlled) switching frequencies, which allow digital signal processing to mask power supply noise.  This technique was used very successfully on Parker Solar Probe.  LuSEE Night requires a large ($\sim$40 kg) battery to provide power through the lunar night.  At sunrise, the system becomes power-positive allowing the battery to recharge, the S-band communication system to send science data back to Earth, and the antenna carousel to be operated.

The LuSEE Night 'Data Controller Board' (DCB) is a computer that interfaces to the other subsystems, including the spectrometer, the S-band communication system (provided by Vulcan Wireless), and the lander's Command and Data Handling (C\&DH) system.  The DCB also controls the motor drive for the antenna carousel.

At the date of this writing (late January 2023), NASA is selecting a CLPS vendor to provide the LuSEE Night launch and landing system.  This mission is designated 'CS-3' and will also carry the European Space Agency (ESA) Lunar Pathfinder (LPF) communications relay satellite into lunar orbit.  LuSEE Night will assist in commissioning of the LPF satellite and will use LPF as its S-band uplink/downlink asset.

\section{Far-field Calibration Source}
NASA has released a Request For Proposals to the CLPS vendors to provide a far-field calibration source (hereafter FFCS) for LuSEE Night (designated as the 'CS-4' mission).  As envisioned, the CS-4 FFCS will be a payload accommodated on another CLPS lunar orbiter or perhaps a dedicated system (e.g. a CubeSat) that carries an instrument that transmits a known pseudo-random waveform that is repeated contiguously with a flux density that corresponds to 10$^{-22}$ to 10$^{-18}$ W/m$^2$/Hz (to $\sim$1\% knowledge) over 10 seconds at the LuSEE Night site on the lunar surface.  This signal corresponds to a frequency comb in the LuSEE Night band (to 51.2 MHz) and the LuSEE Night spectrometer will correlate against this signal as CS-4 passes overhead horizon-to-horizon.  The CS-4 FFCS will be required to provide at least 30 passes over LuSEE Night and to operate for 50 Earth days and then cease operation.  This FFCS will provide LuSEE Night a known signal over a range of altitudes and azimuths that will be used to calibrate the antenna pattern and system voltage response and chromaticity.

\section{Landing Site}

\begin{figure}[htbp]
  \centering
  \includegraphics[width=82mm]{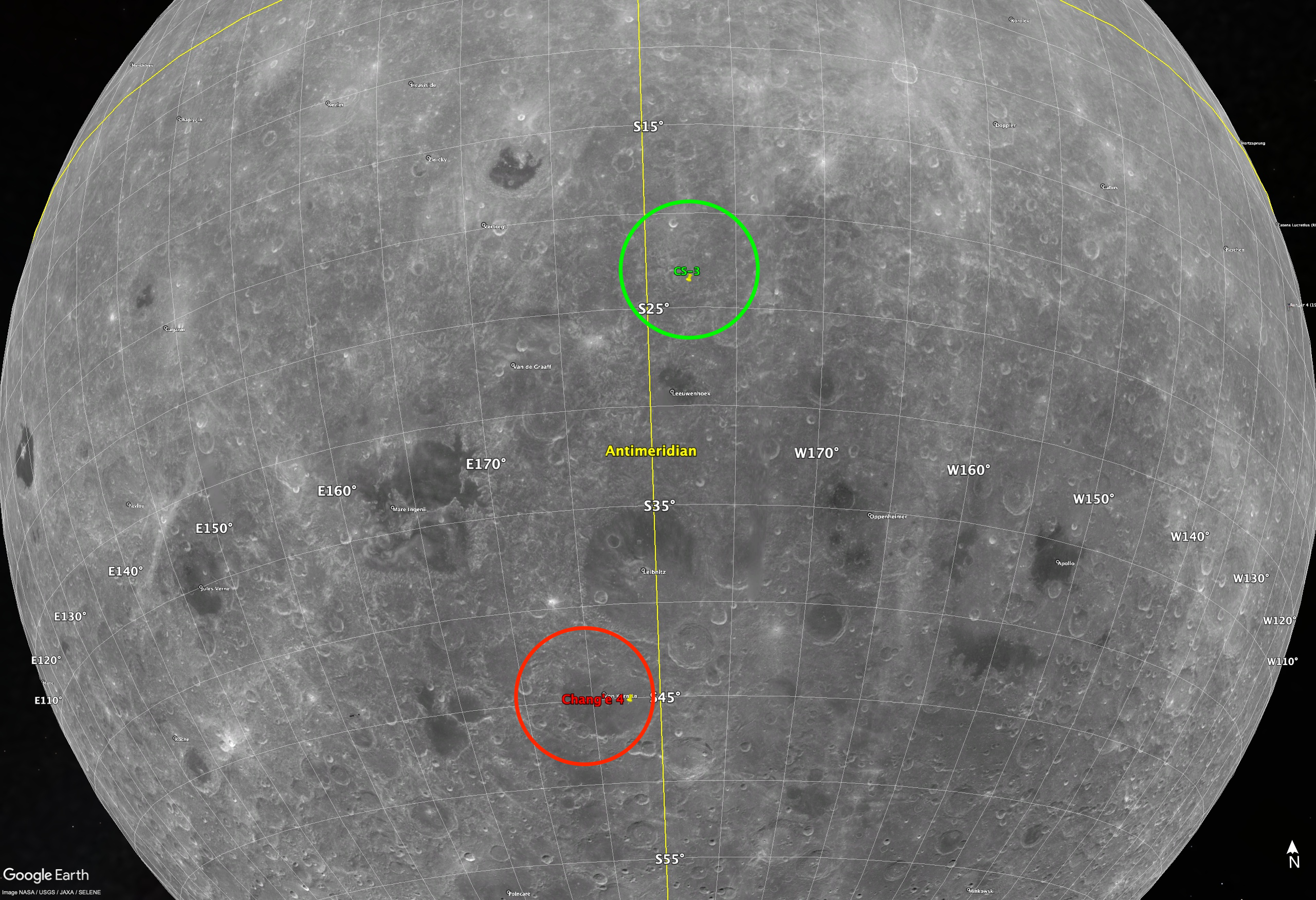}
  \caption{The LuSEE Night landing site (within green circle) on the lunar farside at  (23.813$^\circ$S, 182.258$^\circ$E).  This site was chosen to minimize terrestrial RFI, provide relatively uniform sky coverage, optimize payload thermal design, and provide favorable downlink with the relay satellite.  The landing site of the Chang'e 4 mission is shown in the red circle and yellow vertical line is the anti-meridian.}
  \label{fig:site}
\end{figure}

The LuSEE Night landing site is a 100m ellipse centered on (23.813$^\circ$S, 182.258$^\circ$E) on the lunar farside near the anti-meridian, shown in Figure \ref{fig:site}.  This site was chosen to be near to the antipode and to therefore minimize terrestrial radio frequency interference \cite{1996GeoRL..23.1287K}, recognizing that at lower frequencies some noise will be refracted by the tenuous lunar ionosphere \cite{2020AdSpR..66.1265B}.  The landing site was also chosen to have a relatively flat horizon.  The aforementioned spatial variations of the galactic power spectrum (brighter plane at high frequencies, brighter poles below 3 MHz) may induce spectral chromaticity as the horizon occults galactic structure and the relatively flat horizon minimizes uncertainties due to this effect.  Finally, the landing site was chosen to have well-mixed regolith to avoid asymmetric dielectric properties below the LuSEE Night antenna system.  The requirement that LuSEE Night operate through the lunar night requires an integrated thermal design, as described above.  To optimize the thermal management, the LuSEE Night radiator requires a view of the dark sky, therefore the requirement to land either North or South of the lunar equator.

\section{Summary}

The LuSEE Night payload is in development and funded by NASA and the DOE.  The system will be delivered and integrated to a lander vehicle for a late 2025 - early 2026 launch and landing on the lunar farside.  LuSEE Night will make unique measurements of the global radio sky below 50 MHz, in the absence of terrestrial and lander EMI, and be a pathfinder for larger, more ambitious lunar radio astronomy projects.

\section*{Acknowledgements}
The LuSEE Night program is funded under NASA Contract 80MSFC22CA018 and DOE Office of Science Contract DE-SC0012704.  We acknowledge the superb technical support of Dr Ahmed Fadl of the NASA Goddard Space Flight Center.

\end{document}